# The GWmodel R package: Further Topics for Exploring Spatial Heterogeneity using Geographically Weighted Models

Binbin Lu[a*], Paul Harris[a], Martin Charlton[a], Chris Brunsdon[b]

a. National Centre for Geocomputation, National University of Ireland Maynooth, Maynooth, Co. Kildare, Ireland

b. School of Environmental Sciences, University of Liverpool, Liverpool, UK

**Abstract**

In this study, we present a collection of local models, termed geographically weighted (GW) models, that can be found within the **GWmodel** R package. A GW model suits situations when spatial data are poorly described by the global form, and for some regions the localised fit provides a better description. The approach uses a moving window weighting technique, where a collection of local models are estimated at target locations. Commonly, model parameters or outputs are mapped so that the nature of spatial heterogeneity can be explored and assessed. In particular, we present case studies using: (i) GW summary statistics and a GW principal components analysis; (ii) advanced GW regression fits and diagnostics; (iii) associated Monte Carlo significance tests for non-stationarity; (iv) a GW discriminant analysis; and (v) enhanced kernel bandwidth selection procedures. General Election data sets from the Republic of Ireland and US are used for demonstration. This study is designed to complement a companion **GWmodel** study, which focuses on basic and robust GW models.

# 1. Introduction

In this study, we present a collection of local (non-stationary) statistical models, termed geographically weighted (GW) models (*1*). A GW model suits situations when spatial data are poorly described by the global (stationary) model form, and for some regions a localised fit provides a better description. This type of approaches uses a moving window weighting technique, where a collection of local models are found at target locations. Commonly, outputs or parameters of a GW model are mapped to provide a useful exploratory tool that can direct a more traditional or sophisticated statistical analysis. For example in a regression context, GW regression (*2-6*) can be used to explore relationships in the data. If relationships are deemed stationary across space, then a basic (non-spatial) regression or a regression that accounts for some spatial autocorrelation effect (e.g. *7*) is sufficient. Conversely, if relationships are deemed non-stationary, the GW regression can be replaced with a sophisticated, spatially-varying coefficient model for improved inference, such as those proposed by Gelfand et al. (*8*) or Assunção (*9*).

Other notable GW models include: GW summary statistics (*1, 10, 11*) ; GW distribution analysis (*12*); GW principal components analysis (GW PCA) (*1, 13*); GW generalised linear models (*1, 14*); GW discriminant analysis (GWDA) (*15*); GW-Geostatistical hybrids (*16-20*); GW methods for outlier detection (*21, 49*); and GW methods for network re-design (*53*). The GW modelling framework continues to evolve (*22*) and GW models have been usefully applied to data from a wide range of disciplines in the natural and social sciences.

Many of the listed GW models are included in an R package **GWmodel** (http://www.r-project.org). Notably, **GWmodel** provides functions to a conduct: (i) a GW PCA; (ii) advanced GW regression fits and diagnostics; (iii) associated Monte Carlo significance tests for non-stationarity; (iv) a GW DA; and (v) enhanced bandwidth selection procedures; where all such functions are utilised in this study. In this respect, our study complements a companion **GWmodel** study (*23*), which focused on basic and robust GW models. The same companion study also presented an advance in addressing collinearity in the GW regression model (following the work of *6, 29, 41, 51, 52*). Our study is structured as follows. Section 2 sets the scene, describing the specification of the weighting matrix in a GW model and the case study data sets (General Election data for the Republic of Ireland and the US). Section 3 describes the use of GW summary statistics and a GW PCA, together with associated Monte Carlo tests. Section 4 describes the fitting of a mixed (semi-parametric) GW regression. Section 5 investigates further topics in GW regression; including: (a) multiple hypothesis tests, (b) collinearity diagnostics and (c) the fitting of heteroskedastic models. Section 6 describes a GW DA. Section 7 describes enhanced bandwidth selection procedures for any GW model. Section 8 concludes this work. At all stages, we provide the R commands so that all of the analyses presented in this study can be reproduced.

## 2. Context, the weighting matrix and the study data sets

*2.1. Context*

GW methods are used to investigate spatial heterogeneity, where the form of the heterogeneity reflects the objective of the under-lying statistic or model. For example, a GW standard deviation (from GW summary statistics) investigates spatial change in data variability; a GW regression investigates spatial change in response and predictor data relationships; a GW variogram (*16*) investigates spatial change in spatial dependence. In all cases, a moving window weighting technique is used, where local models are calibrated at (sampled or un-sampled) locations (i.e. the window's centre). For an individual model at some calibration location, all neighbouring observations are weighted according to the properties of a distance-decay kernel function, and the model is locally-fitted to this weighted data. Thus the geographical weighting solely applies to the data in all GW methods, where each local model is fitted to its own GW data (sub-) set. The size of the window over which this localised model might apply is controlled by the kernel function's bandwidth. Small bandwidths lead to more rapid spatial variation in the results, while large bandwidths yield results increasingly close to the global model solution. The GW modelling paradigm encompasses many methods; each locally-adapted from a global form.

## 2.2. Building the weighting matrix

Key to GW modelling is the weighting matrix, which sets the spatial dependency in the data. Here $\mathbf{W}(u_i, v_i)$ is a $n \times n$ diagonal matrix (where $n$ is the sample size) denoting the geographical weighting attached to each observation point, for any model calibration point $i$ at location $(u_i, v_i)$. Thus a different weighting matrix is found at each calibration point. This matrix is determined according to the following three key elements: (i) the type of distance metric; (ii) the type of kernel weighting function; and (iii) the kernel weighting function's bandwidth. For the distance metric, **GWmodel** permits the Minkowski family of distance metrics, where the power of the Minkowski distance $p$, needs to be specified. For example, if $p = 2$ (the default), then the usual Euclidean distance metric results; or if $p = 1$, then the Manhattan (or Taxi cab) distance metric results. As an example, Lu et al. (*24*) fit GW regressions using Euclidean and non-Euclidean distance metrics.

Four of the five kernel weighting functions available in **GWmodel** are defined in Table 1. Each function includes the bandwidth parameter (**r** or **b**), which controls the rate of decay. All functions are defined in terms of weighting the sample data, where **j** is the index of the observation point and $\boldsymbol{d}_{ij}$ the distance between the points indexed by **i** and **j**.

For the box-car and bi-square functions, the bandwidth $r$ can be specified beforehand (i.e. a fixed distance) or specified as the distance between the point $i$ and its $N^{th}$ nearest neighbour, where $N$ is specified beforehand (i.e. an adaptive distance). The bi-square function gives fractional decaying weights according to the proximity of the data to each point $i$, up until a fixed distance or a distance according to a specified $N^{th}$ nearest neighbour. The local search strategy for this and the box-car function is simply $N$ neighbours within a fixed radius $r$ or $N$ nearest neighbours for an adaptive approach. Both functions can suffer from discontinuity, although the bi-square function can be defined with a bandwidth that uses all of the data to minimise such problems.

The Gaussian and exponential functions are continuous and use all the data. Their weights decay according to a Gaussian or exponential curve. According to the bandwidth set, data that are a long way from the point $i$ receive virtually zero weight. The key difference between these functions is their behaviour at the origin. Usually these continuous functions are defined with a fixed bandwidth $b$, but can be constructed to behave in an adaptive manner. The bi-square function is useful as it can provide an intermediate weighting between the box-car and the Gaussian functions. To get similar weights from the bi-square and Gaussian functions, the bandwidths $r$ and $b$ can be approximately related by $r \cong \left(3\sqrt{2}/2\right)b$. For all functions, if $r$ or $b$ is set suitably large enough, then all of the data can receive a weight of one and the corresponding global model or statistic is found.

Table 1 Four kernel weighting functions.

| | | |
|---|---|---|
| ***Box-car*** | $w_{ij} = 1$ if $d_{ij} \leq r$ | $w_{ij} = 0$ otherwise |
| ***Bi-square*** | $w_{ij} = \left(1 - \left(d_{ij}/r\right)^2\right)^2$ if $d_{ij} \leq r$ | $w_{ij} = 0$ otherwise |
| ***Gaussian*** | $w_{ij} = \exp\left(-d_{ij}^2/2b^2\right)$ | |
| ***Exponential*** | $w_{ij} = \exp\left(-d_{ij}/b\right)$ | |

Bandwidths can be: (a) user-specified, when there exists some strong prior belief to do so; (b) optimally- (or automatically-) specified using cross-validation and related approaches, provided there exists an objective function (i.e. the method can be used as a predictor); or (c) user-specified, but guided by (b) where an automated approach is not viewed as a panacea for bandwidth selection (*25*). In **GWmodel**, automated bandwidths can be found for GW regression, GW regression with a locally compensated ridge term (to address local collinearity problems), generalised GW regression, GW DA and GW PCA.

*2.3. Study data sets*

**GWmodel** comes with five example data sets, these are: (1) `Georgia`, (2) `LondonHP`, (3) `DubVoter`, (4) `EWHP` and (5) `USelect`. For this article's presentation of GW models, we use as case studies, the `DubVoter` and `USelect` data sets.

*2.3.1. Dublin 2004 voter turnout data*

The `DubVoter` data is the main study data set and is used throughout sections 3 to 5 and section 7. This data is composed of nine percentage variables[1], measuring: (A) voter turnout in the Irish 2004 Dáil elections and (B) eight characteristics of social structure (census data); for $n$ = 322 Electoral Divisions (EDs) of Greater Dublin. Kavanagh et al. (*26*) modelled this data using GW regression; with voter turnout (*GenEl2004*), the dependent variable (i.e. the percentage of the population in each ED who voted in the election). The eight independent variables measure the percentage of the population in each ED, with respect to:

- one year migrants (i.e. moved to a different address one year ago) (*DiffAdd*);
- local authority renters (*LARent*);

---
[1] Observe that none of these variables constitute a closed system and as such, do not need to be treated as compositional data.

- social class one (high social class) (*SC1*);

- unemployed (*Unempl*);

- without any formal educational (*LowEduc*);

- age group 18-24 (*Age18_24*);

- age group 25-44 (*Age25_44*); and

- age group 45-64 (*Age45_64*).

The independent variables reflect measures of migration, public housing, high social class, unemployment, educational attainment, and three broad adult age groups. Other GW model studies using versions of this data include that of Harris et al. (*13*) and Gollini et al. (*23*).

*2.3.2. US 2004 election data*

The `USelect` data is only used in section 6, for demonstrating a GW DA. It consists of the results of the 2004 US presidential election at the county level ($n = 3111$), together with a collection of socio-economic (census) variables (*27*). A variant of this data has been used for the visualisation of GW DA outputs in Foley and Demšar (*28*). In terms of the election results, Bush or Kerry was always the winner within a county; while in some counties, the supporting ratio for a candidate ranged from 45% to 55%, which for our purposes is viewed as an 'unclear-winner' or a 'borderline' result. Thus for our version of this data set, we produce a categorical dependent variable with three classes: (i) Bush winner, (ii) Kerry winner and (iii) Borderline. If we proceed with just two classes: (a) Bush winner and (b) Kerry winner; then an issue arises in that a GW logistic regression may provide a simpler approach to the local modelling of this data, than that found with a GW DA (*55*) (although observe that both methods can be applied to categorical dependent data with more than two classes (*43*)). For the `USelect` data, the five independent variables are taken the same as that used in Foley and Demšar (*28*), as follows:

- percentage unemployed (*unemployed*)
- percentage of adults over 25 with 4 or more years of college education (*pctcoled*)
- percentage of persons over the age of 65 (*PEROVER65*)
- percentage urban (*pcturban*)
- percentage white (*WHITE*).

## 3. Exploration with GW summary statistics and GW PCA

This first section on GW modelling presents case studies on the use of GW summary statistics and a GW PCA. For demonstration, we investigate the `DubVoter` data.

*3.1. GW summary statistics*

Although simple to calculate and map, GW summary statistics can act as a vital precursor to an application of a subsequent GW model. For example, GW standard deviations will highlight areas of high variability for a given variable; areas where a subsequent application of a GW PCA or a GW regression may warrant close scrutiny. For attributes $x$ and $y$ at any location $i$ where $w_{ij}$ accords to a kernel function of section 2, definitions for a GW mean, GW standard deviation and GW correlation coefficient are respectively

$$\mu(x_i) = \sum_{j=1}^{n} w_{ij} x_j \bigg/ \sum_{j=1}^{n} w_{ij} \qquad (1)$$

$$s(x_i) = \sqrt{\sum_{j=1}^{n} w_{ij} (x_j - \mu(x_i))^2 \bigg/ \sum_{j=1}^{n} w_{ij}} \qquad (2)$$

and

$$\rho(x_i, y_i) = c(x_i, y_i) / (s(x_i) s(y_i)) \quad (3)$$

with the GW covariance

$$c(x_i, y_i) = \sum_{j=1}^{n} w_{ij} \{(x_j - \mu(x_i))(y_j - \mu(y_i))\} \bigg/ \sum_{j=1}^{n} w_{ij} . \quad (4)$$

*3.2. GW PCA*

In a PCA, a set of $m$ correlated variables are transformed in to a new set of $m$ uncorrelated variables called components. Components are linear combinations of the original variables that can allow for a better understanding of sources of variation and trends in the data. Its use as a dimension reduction technique is viable if the first few components account for most of the variation in the original data. In a GW PCA, a series of localised PCAs are computed, where the local component outputs (variances, loadings and scores) are mapped, permitting a local identification of any change in structure of the multivariate data. This local exploration can pinpoint locations where results from a PCA are inappropriate. This in turn, allows for better-informed model decisions for any analysis that may follow, such as a clustering or regression analysis when orthogonal input data are required. GW PCA can assess: (i) how (effective) data dimensionality varies spatially and (ii) how the original variables influence each spatially-varying component.

More formally, if an observation location $j$ has coordinates $(u,v)$, then a GW PCA involves regarding a vector of observed variables $\mathbf{x}_j$ as having a certain dependence on its location $j$, where $\mathbf{\mu}(u,v)$ and $\mathbf{\Sigma}(u,v)$ are the local mean vector and the local variance-covariance matrix, respectively. The local variance-covariance matrix is

$$\mathbf{\Sigma}(u,v) = \mathbf{X}^T \mathbf{W}(u,v) \mathbf{X} \quad (5)$$

where **X** is the $n \times m$ data matrix; and $\mathbf{W}(u,v)$ is a diagonal matrix of geographic weights, generated by a kernel function of section 2. The kernel's bandwidth can be user-specified or found optimally via cross-validation (*13*). To find the local principal components at location $(u_j, v_j)$, the decomposition of the local variance-covariance matrix provides the local eigenvalues and local eigenvectors (or loading vectors) with

$$\mathbf{L}(u_j, v_j) \mathbf{V}(u_j, v_j) \mathbf{L}(u_j, v_j)^{\mathrm{T}} = \mathbf{\Sigma}(u_j, v_j) \tag{6}$$

where $\mathbf{L}(u_j, v_j)$ is a matrix of local eigenvectors; $\mathbf{V}(u_j, v_j)$ is a diagonal matrix of local eigenvalues; and $\mathbf{\Sigma}(u_j, v_j)$ is the local covariance matrix. A matrix of local component scores $\mathbf{T}(u_j, v_j)$ can be found using

$$\mathbf{T}(u_j, v_j) = \mathbf{X}\mathbf{L}(u_j, v_j). \tag{7}$$

If we divide each local eigenvalue by $\mathrm{tr}(\mathbf{V}(u_j, v_j))$, then we find the local proportion of the total variance (PTV) in the original data accounted for by each component. Thus at each location *j* for a GW PCA with *m* variables, there are *m* components, *m* eigenvalues, *m* sets of component loadings (with each set $m \times m$), and *m* sets of component scores (with each set $n \times m$). We can obtain eigenvalues and their associated eigenvectors at un-observed locations, but as no data exists for these locations, we cannot obtain component scores.

### 3.3. Monte Carlo tests for non-stationarity

For GW summary statistics and GW PCA, Monte Carlo tests are possible that test for non-stationarity (*1, 13*). Tests confirm whether or not the GW summary statistic or aspects of the GW PCA are *significantly* different to that found by chance or artefacts of random variation in the data. Here the sample data are successively randomised and the GW model is applied after each randomisation. A basis of a significance test is then possible by comparing the true result with results from a large number of randomised distributions. The randomisation hypothesis is that any pattern seen in the data occurs by chance and therefore any permutation of the data is equally likely.

As an example for GW correlation, the test proceeds as follows: (i) calculate the true GW correlation at all locations; (ii) randomly choose a permutation of the data where the coordinates are kept in the same pairs, as are the chosen attribute pairs; (iii) calculate a simulated GW correlation at all locations using the randomised data of (ii); (iv) repeat steps (ii) and (iii), say 99 times; (v) at each location $i$, rank the one true GW correlation with the 99 simulated GW correlations; (vi) at each location $i$, if the true GW correlation lies in the top or bottom 2.5% tail of the ranked distribution then the true GW correlation can be said to be significantly different (at the 95% level) to such a GW correlation found by chance. The results from this Monte Carlo test are mapped.

For a GW PCA, a similar procedure is followed where the test evaluates whether the local eigenvalues vary significantly across space. Here the paired coordinates are successively randomised amongst the variable data set and after each randomisation, GW PCA is applied (with an optimally re-estimated bandwidth) and the standard deviation (SD) of a given local eigenvalue is calculated. The true SD of the same local eigenvalue is then included in a ranked distribution of SDs. Its position in this ranked distribution relates to whether there is significant (spatial) variation in the chosen local eigenvalue. The results from this Monte Carlo test are presented via a graph.

*3.4. Examples: GW correlations*

For a demonstration of an analysis using a GW summary statistic, we calculate GW correlations to investigate the local relationships between: (a) voter turnout (*GenEl4004*) and *LARent* and (b) *LARent* and *Unempl*. In the former case, the correlations provide a preliminary assessment of relationship non-stationarity between the dependent and an independent variable of a GW regression of sections 4 and 5. In the latter case, the correlations provide an assessment of local collinearity between two independent variables of such a GW regression (*29*). In both cases, we specify a bi-square kernel. Furthermore, as the spatial arrangement of the EDs in Greater Dublin is not a tessellation of equally sized zones, it makes sense to specify an adaptive bandwidth, that we user-specify at $N = 48$. This entails that the bi-square kernel will change in radius, but will always include the closest 48 EDs for each local correlation. Bandwidths for GW correlations cannot be found optimally using cross-validation (although see *53*, for an alternative). We also conduct the corresponding Monte Carlo tests for the two GW correlation specifications. Commands to conduct our GW correlation analysis are as follows, where we use the function *gwss* to find GW summary statistics and **montecarlo.gwss** to conduct the Monte Carlo tests. Commands include those to visualise the outputs (Figure 1).

```
R > library(GWmodel)
R > library(RColorBrewer)
R > data(DubVoter)

R > gwss.1 <- gwss(Dub.voter,vars = c("GenEl2004", "LARent", "Unempl"),
kernel="bisquare", adaptive=TRUE, bw=48)

R > gwss.mc <- montecarlo.gwss(Dub.voter,vars = c("GenEl2004", "LARent",
"Unempl"), kernel="bisquare", adaptive=TRUE, bw=48)

R > gwss.mc.data <- data.frame(gwss.mc)
R > gwss.mc.out.1 <-ifelse(gwss.mc.data$Corr_GenEl2004.LARent < 0.975 &
R > gwss.mc.data$Corr_GenEl2004.LARent > 0.025 , 0, 1)
R > gwss.mc.out.2 <-ifelse(gwss.mc.data$Corr_LARent.Unempl < 0.975 &
R > gwss.mc.data$Corr_LARent.Unempl > 0.025 , 0, 1)
R > gwss.mc.out <- data.frame(Dub.voter$X, Dub.voter$Y, gwss.mc.out.1,
gwss.mc.out.2)

R > gwss.mc.out.1.sig <- subset(gwss.mc.out, gwss.mc.out.1==1, select =
c(Dub.voter.X, Dub.voter.Y, gwss.mc.out.1))
R > gwss.mc.out.2.sig <- subset(gwss.mc.out, gwss.mc.out.2==1, select =
c(Dub.voter.X, Dub.voter.Y, gwss.mc.out.2))
R > pts.1 <- list("sp.points", cbind(gwss.mc.out.1.sig[,1],
gwss.mc.out.1.sig[,2]), cex=2, pch="+", col="black")
R > pts.2 <- list("sp.points", cbind(gwss.mc.out.2.sig[,1],
gwss.mc.out.2.sig[,2]), cex=2, pch="+", col="black")
```

```
R > mypalette.gwss.1 <-brewer.pal(5,"Blues")
R > mypalette.gwss.2 <-brewer.pal(6,"Greens")

R > map.na <- list("SpatialPolygonsRescale", layout.north.arrow(),
offset = c(329000,261500), scale = 4000, col=1)
R > map.scale.1 <- list("SpatialPolygonsRescale", layout.scale.bar(),
offset = c(326500,217000), scale = 5000, col=1, fill =
c("transparent", "green"))
R > map.scale.2  <- list("sp.text", c(326500,217900), "0", cex=0.9,
col=1)
R > map.scale.3  <- list("sp.text", c(331500,217900),"5km",
cex=0.9,col=1)
R > map.layout.1 <-list(map.na,map.scale.1,map.scale.2,map.scale.3,pts.1)
map.layout.2 <- list(map.na,map.scale.1,map.scale.2,map.scale.3,pts.2)

R > X11(width=10,height=12)
R > spplot(gwss.1$SDF,"Corr_GenEl2004.LARent",key.space = "right",
col.regions = mypalette.gwss.1,at=c(-1,-0.8,-0.6,-0.4,-0.2,0),
par.settings = list(fontsize=list(text=15)), main = list(label="GW
correlations: GenEl2004 and LARent", cex=1.25), sub=list(label="+
Results of Monte Carlo test", cex=1.15), sp.layout=map.layout.1)

R > X11(width=10,height=12)
R > spplot(gwss.1$SDF,"Corr_LARent.Unempl",key.space = "right",
col.regions=mypalette.gwss.2,at=c(-0.2,0,0.2,0.4,0.6,0.8,1),
par.settings=list(fontsize=list(text=15)), main=list(label="GW
correlations: LARent and Unempl", cex=1.25), sub=list(label="+
Results of Monte Carlo test", cex=1.15), sp.layout=map.layout.2)
```

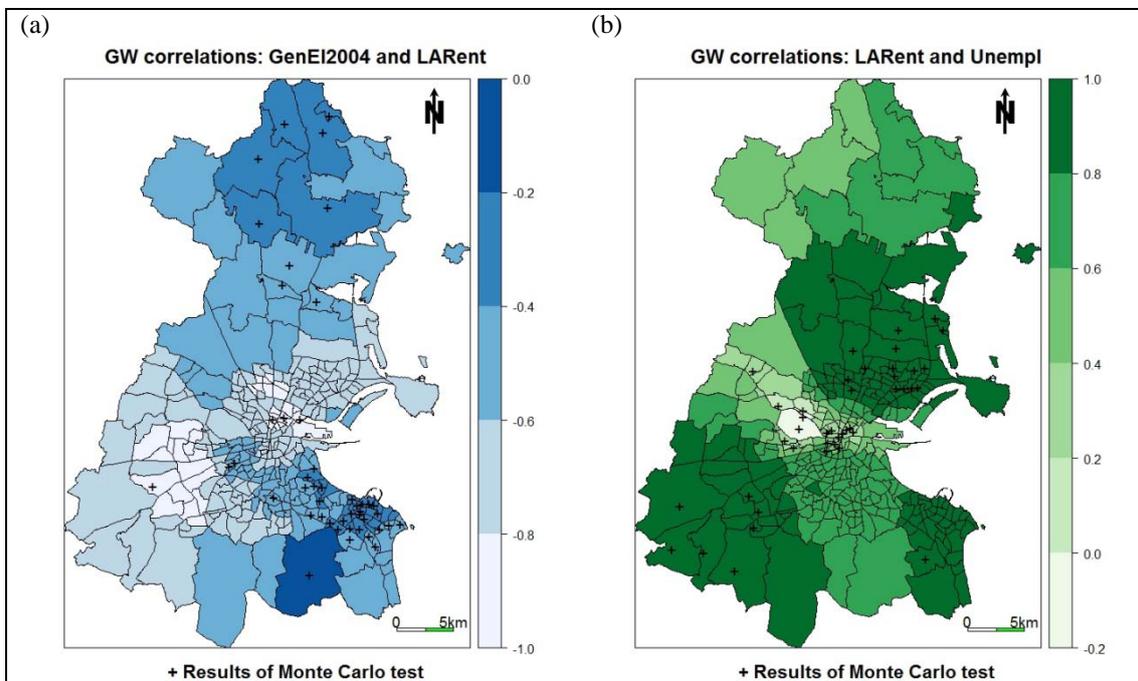

Figure 1 GW correlations and associated Monte Carlo tests for: (a) *GenEl2004* and *LARent*; and (b) *LARent* and *Unempl*. Global correlations are -0.68 and 0.67, respectively.

From Figure 1(a), the relationship between turnout and LARent appears non-stationary, where this relationship is strongest in areas of central and south-west Dublin. Here turnout tends to be low, while local authority renting tends to be high. The associated Monte Carlo test suggests many instances of unusual relationships, such as those found in the north that are unusually weak. From Figure 1(b), consistently strong positive correlations between *LARent* and *Unempl* are found in three distinct areas of Greater Dublin; areas where local collinearity effects in the GW regression of sections 4 and 5 are likely to be a cause for concern (see also *23*).

### *3.5. Examples: PCA to GW PCA*

For applications of PCA and GW PCA, we investigate these eight variables: *DiffAdd*, *LARent*, *SC1*, *Unempl*, *LowEduc*, *Age18_24*, *Age25_44* and *Age45_64*. We standardise the data and specify the PCA with the covariance matrix. The same (globally) standardised data is also used in the GW PCA calibration, which is similarly specified with (local) covariance matrices. The effect of this standardisation is to make each variable have equal importance in the subsequent analysis (at least for the PCA case)[2]. The PCA results (PTV data and loadings) are found using *scale* and *princomp* functions, as follows:

```
R > Data.scaled <- scale(as.matrix(Dub.voter@data[,4:11]))
R > pca <- princomp(Data.scaled, cor=F)
R > (pca$sdev^2/sum(pca$sdev^2))*100

Comp.1 Comp.2 Comp.3 Comp.4  Comp.5 Comp.6 Comp.7 Comp.8
36.084 25.586 11.919 10.530   6.890  3.679  3.111  2.196

R > pca$loadings
```

---

[2] The use of un-standardised data, or the use of locally-standardised data with GW PCA is a subject of current research.

```
Loadings:
         Comp.1 Comp.2 Comp.3 Comp.4 Comp.5 Comp.6 Comp.7 Comp.8
DiffAdd  -0.389 -0.444        -0.149  0.123  0.293  0.445  0.575
LARent   -0.441  0.226  0.144  0.172  0.612  0.149 -0.539  0.132
SC1       0.130 -0.576        -0.135  0.590 -0.343         -0.401
Unempl   -0.361  0.462         0.189  0.197         0.670 -0.355
LowEduc  -0.131  0.308 -0.362 -0.861
Age18_24 -0.237         0.845 -0.359 -0.224                -0.200
Age25_44 -0.436 -0.302 -0.317        -0.291  0.448 -0.177 -0.546
Age45_64  0.493  0.118  0.179 -0.144  0.289  0.748  0.142 -0.164
```

From the PTV data, the first two components collectively account for 61.6% of the variation in the data. From the loadings, components one and two mainly represents older (*Age45_64*) and affluent (*SC1*) residents, respectively. However, these results may not reliably represent local social structure, and an application of GW PCA may be useful. Here a bandwidth for GW PCA is found using cross-validation, where it is necessary to decide *a priori* on the number of components, $k$ to retain, provided $m \neq k$. Thus we choose to find an optimal adaptive bandwidth using a bi-square kernel, with $k = 3$. Here the ***bw.gwpca*** function is used within the following set of commands:

```
R > Coords <- as.matrix(cbind(Dub.voter$X,Dub.voter$Y))
R > Data.scaled.spdf <- SpatialPointsDataFrame(Coords,
as.data.frame(Data.scaled))
R > bw.gwpca.1 <-bw.gwpca(Data.scaled.spdf,vars =
colnames(Data.scaled.spdf@data), k=3, adaptive=TRUE)
```

Inspecting the ***bw.gwpca.1*** object indicates a bandwidth of $N = 131$ will be used to calibrate the GW PCA fit. Observe that we now specify all $k = 8$ components, but we will focus our investigation on only the first two components. This specification ensures that the PTV data is estimated correctly. The GW PCA fit is conducted using the ***gwpca*** function:

```
R > gwpca.1 <- gwpca(Data.scaled.spdf, vars =
colnames(Data.scaled.spdf@data), bw=bw.gwpca.1, k=8, adaptive=TRUE)
```

The GW PCA outputs are visualised and interpreted, focusing on: (1) how data dimensionality varies spatially and (2) how the original variables influence the components.

For the former, the spatial distribution of local PTV for the first two components can be mapped. For the latter, we look at the change in size and sign of the eight local loadings together, for a given component, at each of the 322 EDs. In this respect, we map multivariate glyphs that have spokes around a central hub in which the length of the spoke corresponds to the size of the local loading, and its colour corresponds to the sign (in this case, blue signifies positive and red signifies negative). The glyphs are scaled relative to the spoke with the largest absolute loading. The variable corresponding to each local loading is always in the same place on the glyph, as follows: *DiffAdd* is at $0^o$ (north); *LARent* is $45^o$ (north-east); *SC1* is $90^o$ (east); *Unempl* is $135^o$ (south-east), *LowEduc* is $180^o$ (south), *Age18_24* is $225^o$ (south-west), *Age25_44* is $270^o$ (west) and *Age45_64* is $315^o$ (north-east). Commands to conduct these visualisations are as follows:

```
R > prop.var <- function(gwpca.obj, n.components) {
return((rowSums(gwpca.obj$var[,1:n.components])/rowSums(gwpca.obj$var
))*100)}
R > var.gwpca <- prop.var(gwpca.1,2)
R > Dub.voter$var.gwpca <- var.gwpca

R > mypalette.gwpca.1 <-brewer.pal(8,"YlGnBu")
R > map.layout.3 <- list(map.na,map.scale.1,map.scale.2,map.scale.3)

R > X11(width=10,height=12)
R > spplot(Dub.voter,"var.gwpca",key.space = "right", col.regions =
mypalette.gwpca.1, cuts=7, par.settings =list(fontsize=list(text=15)),
main=list(label="GW PCA: PTV for local components 1 to 2", cex=1.25),
sp.layout=map.layout.3)

R > loadings.1 <- gwpca.1$loadings[,,1]

R > X11(width=10,height=12)
R > plot(Dub.voter)
glyph.plot(loadings.1,Coords,r1=20,add=T,alpha=0.85)
title(main=list("GW PCA: Multivariate glyphs of loadings", cex=1.75,
col="black", font=1), sub=list("For component 1", cex=1.5,
col="black", font=3))
```

Figure 2(a) presents the local PTV map, where there is clear spatial variation in the PTV data. Higher percentages are located in the south and in central Dublin, whilst lower percentages are located in the north. The PTV data is also generally higher in the local case, than in the global case (at 61.6%). Figure 2(b) presents a multivariate glyph map for the loadings on the first component, where a spatial preponderance of glyphs of one colour or another, or larger spokes on the same variables provide a general indication of the structures being represented at each of the 322 EDs. This map is not intended to be scrutinised in detail, but clearly indicates strong local trends in social structure across Greater Dublin.

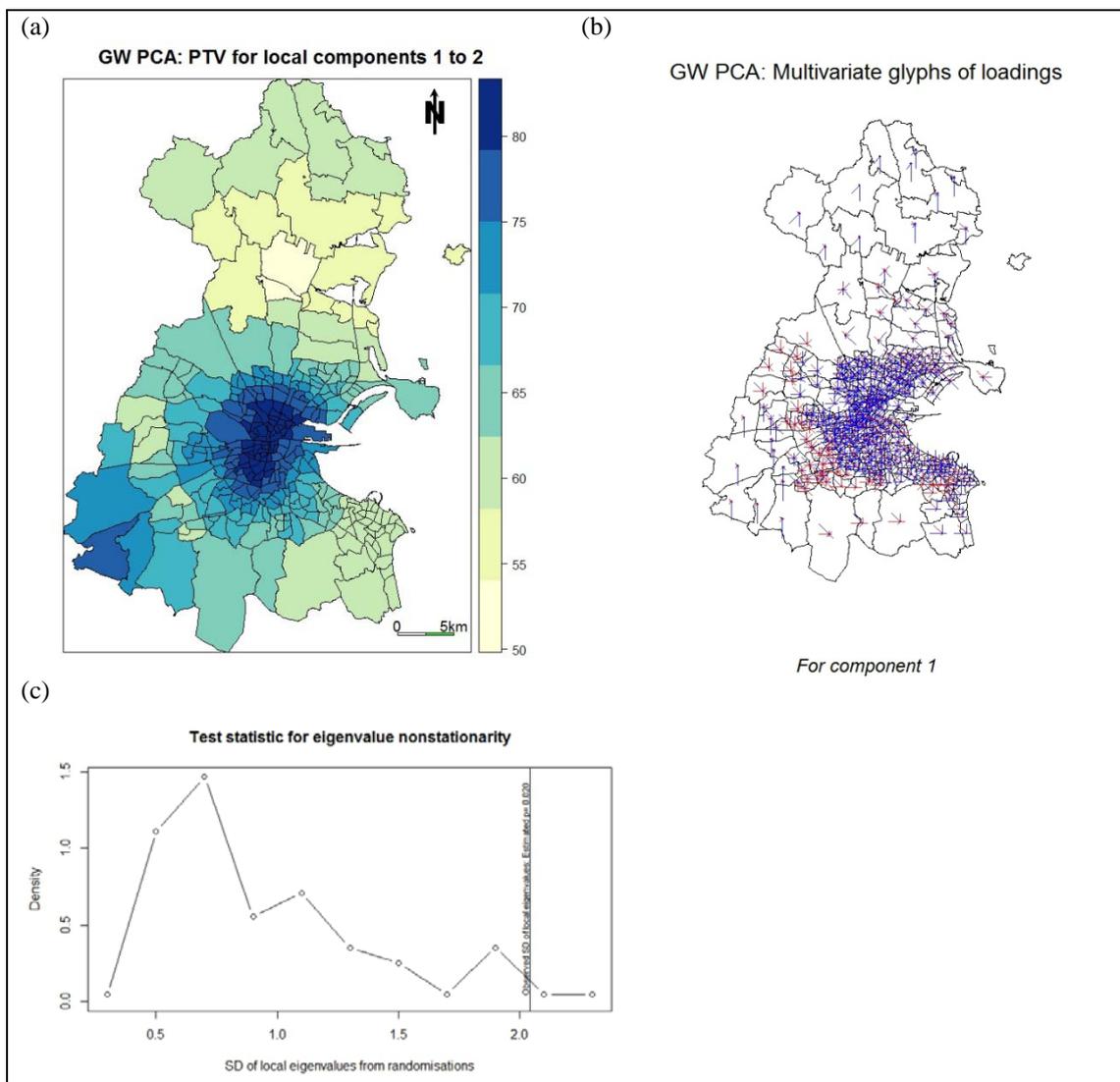

Figure 2 GW PCA results: (a) PTV data for the first two components; (b) multivariate glyphs of the loadings for all eight components; and (c) Monte Carlo test for first component only.

To provide support for our chosen GW PCA specification, the associated Monte Carlo test evaluates whether the local eigenvalues for the first component vary significantly across space. The results are given in Figure 2(c), where the *p*-value for the true SD of the eigenvalues is calculated at 0.02. Thus an application of GW PCA is considered worthy as the null hypothesis of local eigenvalue stationarity is firmly rejected at the 95% level, for the dominant first component. Commands to conduct this test are as follows:

```
R > gwpca.mc <-montecarlo.gwpca.2(Data.scaled.spdf, vars =
colnames(Data.scaled.spdf@data), k=3, adaptive=TRUE)

R > X11(width=8,height=5)
R > plot.mcsims(gwpca.mc)
```

## 4. Mixed model building for GW regression

The first and most commonly applied GW model is GW regression (*3, 4*). This model enables an exploration of spatially-varying data relationships, via the visualisation and interpretation of sets of local regression coefficients and associated estimates. For **GWmodel,** a range of GW regression models are included: (i) basic; (ii) robust; (iii) generalised; (iv) mixed; (v) heteroskedastic and (vi) locally compensated ridge. The use of basic, robust and locally compensated ridge GW regression functions is presented in the companion paper of Gollini et al. (*23*). For this paper, we present the use of basic and mixed GW regression functions; together with the further GW regression topics of section 5.

### *4.1. Basic and mixed GW regression*

The basic form of the GW regression model is

$$y_i = \beta_{i0} + \sum_{k=1}^{m} \beta_{ik} x_{ik} + \varepsilon_i \qquad (8)$$

where $y_i$ is the dependent variable at location *i*; $x_{ik}$ is the value of the *k*th independent variable at location *i*; *m* is the number of independent variables; $\beta_{i0}$ is the intercept parameter at location $i$; $\beta_{ik}$ is the local regression coefficient for the $k$th independent variable at location $i$; and $\varepsilon_i$ is the random error at location $i$. A key assumption for this basic (and related forms of) GW regression is that the local coefficients vary at the same scale and rate across space (depending on the particular kernel weighting function that is specified). However, some coefficients (and relationships) may be expected to have different degrees of variation over the study region. In particular, some coefficients (and relationships) are viewed as constant (or stationary) in nature, whilst others are not. For these situations, a mixed GW regression can be specified (*1, 2*). This semi-parametric model treats some coefficients as global (and stationary), whilst the rest are treated as local (and non-stationary), but with the same rate of spatial variation. This model's general form can be written as

$$y_i = \sum_{j=1,k_a} a_j x_{ij}(a) + \sum_{l=1,k_b} b_l(u_i,v_i) x_{il}(b) + \varepsilon_i \qquad (9)$$

where $\{a_1,......,a_{k_a}\}$ are the $k_a$ global coefficients; $\{b_1(u_i,v_i),......,b_{k_b}(u_i,v_i)\}$ are the $k_b$ local coefficients; $\{x_{i1}(a),......,x_{ik_a}(a)\}$ are the independent variables associated with global coefficients; and $\{x_{i1}(b),......,x_{ik_b}(b)\}$ are the independent variables associated with local coefficients. In a vector-matrix notation, equation (9) can be rewritten as

$$y = X_a \mathbf{a} + X_b \mathbf{b} + \varepsilon \qquad (10)$$

where *y* is the vector of dependent variables; *$X_a$* is the matrix of globally-fixed variables; *a* is the vector of *$k_a$* global coefficients; *$X_b$* is the matrix of locally-varying variables; and *b* is the matrix of local coefficients. To calibrate this model in **GWmodel,** we follow that of Brunsdon et al. (*2*), where a back-fitting procedure is adopted (*30*). If we define the hat matrix for the global regression part of the model, as *$S_a$*; and that for the GW regression part, as *$S_b$*; then equation (9) can be rewritten as

$$\widehat{y} = \widehat{y}_a + \widehat{y}_b \tag{11}$$

where the two components, $\widehat{y}_a$ and $\widehat{y}_b$ can be expressed as

$$\widehat{y}_a = S_a y \tag{12}$$

$$\widehat{y}_b = S_b y \tag{13}$$

and thus the calibration procedure can be briefly described in the following six steps:

Step 1. Supply an initial value for $\widehat{y}_a$, say $\widehat{y}_a^{(0)}$, practically by regressing **$X_a$** on **y** using ordinary least squares (OLS).

Step 2. Set $i=1$.

Step 3. Set $\widehat{y}_b^{(i)} = S_b \left[ y - \widehat{y}_a^{(i-1)} \right]$.

Step 4. Set $\widehat{y}_a^{(i)} = S_a \left[ y - \widehat{y}_b^{(i)} \right]$.

Step 5. Set $i=i+1$.

Step 6. Return to Step 3 unless $\hat{y}^{(i)} = \hat{y}_a^{(i)} + \hat{y}_b^{(i)}$ converges to $\hat{y}^{(i-1)}$.

An alternative mixed GW regression fitting procedure is given in Fotheringham et al. (*1*), that uses a method from Speckman (*31*). This alternative is not as computationally intensive as that presented here, and is under consideration for future mixed model in **GWmodel.**

*4.2. Monte Carlo tests for regression coefficient non-stationarity*

For a mixed GW regression, difficulties arise when deciding whether a relationship should be fixed globally or allowed to vary locally. Here Fotheringham et al. (*32*) adopt a stepwise procedure, where all possible combinations of global and locally-varying relationships are tested, and an optimal mixed model is chosen according to a minimised AIC value. This approach is comprehensive, but computationally expensive, and is utilised in the GW regression 4.0 executable software (*33*). Alternatively, a Monte Carlo approach can be used to test for significant (spatial) variation in each regression coefficient (or relationship) from the basic GW regression model, where the null hypothesis is that the relationship between dependent and independent variable is constant (*1, 3, 4*). The procedure is analogous to that presented for the local eigenvalues of a GW PCA in section 3.3, where for the basic GW regression the true variability in each local regression coefficient is compared to that found from a series of randomised data sets. If the true variance of the coefficient does not lie in the top 5% tail of the ranked results, then the null hypothesis can be accepted at the 95% level; and the corresponding relationship should be globally-fixed when specifying the mixed GW regression. Observe, that if all relationships are viewed as non-stationary, then the basic GW regression should be preferred. Conversely, if all relationships are viewed as stationary, then the standard global regression should be preferred. Advances on the mixed GW regression model, where the relationships can be allowed to vary at different rates across space can be found in Yang et al. (*50*).

### 4.3. Example: mixed GW regression model specification

We now demonstrate modelling building for mixed GW regression using the `DubVoter` data. First we calibrate a basic GW regression. We then conduct the Monte Carlo test on this model's outputs, to gauge for significant variation (or non-stationarity) in each coefficient, including the intercept term. Finally, we fit a mixed GW regression according to the results of the Monte Carlo test. Our regressions investigate the local/global relationships between the response: *GenEl2004* and these eight predictors: *DiffAdd*, *LARent*, *SC1*, *Unempl*, *LowEduc*, *Age18_24*, *Age25_44* and *Age45_64*. For both GW regressions, we specify a bi-square kernel with an adaptive bandwidth.

Table 2 Monte Carlo test for the basic GW regression.

| Variable | Intercept | DiffAdd | LARent | SC1 | Unempl |
|---|---|---|---|---|---|
| *p*-value | 0.35 | 0.17 | 0.28 | 0.02 | 0.00 |
| **Variable** | *LowEduc* | *Age18_24* | *Age25_44* | *Age45_64* | |
| *p*-value | 0.19 | 0.04 | 0.29 | 0.19 | |

The optimal bandwidth for the basic GW regression is found at *N = 109* in accordance to an automatic AICc approach via the function **bw.gwr**. This bandwidth is then used calibrate the basic GW regression via the function **gwr.basic**. We then conduct the Monte Carlo test where the results are presented in Table 2. The results suggest (say, at the 95% level) that the *Intercept* term together with the *DiffAdd*, *LARent*, *LowEduc*, *Age25_44* and *Age45_64* variables, should all be fixed as global in the mixed model. Accordingly, the mixed model is calibrated using the function **gwr.mixed** with the same adaptive bandwidth as that found for the basic model. Observe that the geographically varying coefficients in the mixed model are less variable than the corresponding coefficients from the basic model, although the same bandwidth is used. Commands to conduct these operations are as follows, where the **print** function imitates the output of the GW regression 3.0 executable software (*34*):

```
R > bw.gwr.1 <- bw.gwr(GenEl2004 ~ DiffAdd + LARent + SC1 + Unempl
+ LowEduc + Age18_24 + Age25_44 + Age45_64, data = Dub.voter,
approach = "AICc",kernel = "bisquare", adaptive = TRUE)
R > bgwr.res <- gwr.basic(GenEl2004 ~ DiffAdd + LARent + SC1 + Unempl +
LowEduc + Age18_24 + Age25_44 + Age45_64, data = Dub.voter,
bw = bw.gwr.1, kernel = "bisquare", adaptive = TRUE)
R > print(bgwr.res)

    *************Summary of GWR coefficient estimates:***************
                  Min.      1st Qu.     Median    3rd Qu.     Max.
    Intercept  53.2300000 73.3200000 81.6600000 95.0700000 116.8000
    DiffAdd    -0.7281000 -0.3338000 -0.1584000  0.1586000   0.5465
    LARent     -0.1949000 -0.1206000 -0.0844400 -0.0369200   0.0940
    SC1        -0.1578000  0.0352800  0.3088000  0.4201000   0.8796
    Unempl     -2.3180000 -1.1440000 -0.7649000 -0.4753000  -0.0925
    LowEduc    -7.6750000 -0.7369000  0.5332000  1.8100000   3.4140
    Age18_24   -0.3970000 -0.2529000 -0.1457000  0.0007642   0.3669
    Age25_44   -1.0950000 -0.7209000 -0.4536000 -0.3048000   0.2184
    Age45_64   -0.9236000 -0.4098000 -0.1102000  0.0467900   0.4931
```

```
R > bgwr.mc <- montecarlo.gwr(GenEl2004 ~ DiffAdd + LARent + SC1 +
Unempl + LowEduc + Age18_24 + Age25_44 + Age45_64, data = Dub.voter,
bw = bw.gwr.1, kernel = "bisquare", adaptive = TRUE)

R > mgwr.res <- gwr.mixed(GenEl2004 ~ DiffAdd + LARent + SC1 + Unempl +
LowEduc + Age18_24 + Age25_44 + Age45_64, data = Dub.voter,
bw = bw.gwr.1, fixed.vars = c("DiffAdd", "LARent", "LowEduc", "Age25_44",
"Age45_64"), intercept.fixed = TRUE, kernel = "bisquare", adaptive =
TRUE)
R > print(mgwr.res)

****Summary of mixed GWR coefficient estimates:*********************
Estimated global variables :   Intercept  DiffAdd   LARent   LowEduc
Age25_44 Age45_64
Estimated global coefficients:   86.31399 -0.15299 -0.11481   0.12894 -
0.53151  -0.2579
Estimated GWR variables :
           Min.    1st Qu.   Median   3rd Qu.    Max.
SC1      0.01948   0.10360   0.19820   0.42970   0.7112
Unempl  -1.03400  -0.77250  -0.65630  -0.52440  -0.0668
Age18_24 -0.40890 -0.20600  -0.12650  -0.06580   0.1139
```

As an example, Figures 3(a-b) present the coefficient surfaces corresponding to *Unempl* found from the basic and mixed GW regressions, respectively. The spatial variation in this coefficient is clearer greater when using the basic GW regression. Differences in the coefficient surfaces primarily occur in the north-west and south-west areas of Dublin. Commands for these maps are as follows:

```
R > mypalette.gwr <- brewer.pal(6, "Spectral")

R > X11(width=10,height=12)
R > spplot(bgwr.res$SDF, "Unempl", key.space = "right", col.regions =
mypalette.gwr, at = c(-3, -2.5, -2, -1.5, -1, -0.5, 0), main = "Basic
GW regression coefficient estimates for Unempl",
sp.layout=map.layout.3)

R > X11(width = 10, height = 12)
R > spplot(mgwr.res$SDF, "Unempl_L", key.space = "right", col.regions =
mypalette.gwr, at = c(-3, -2.5, -2, -1.5, -1, -0.5, 0), main = "Mixed
GW regression coefficient estimates for Unempl",
sp.layout=map.layout.3)
```

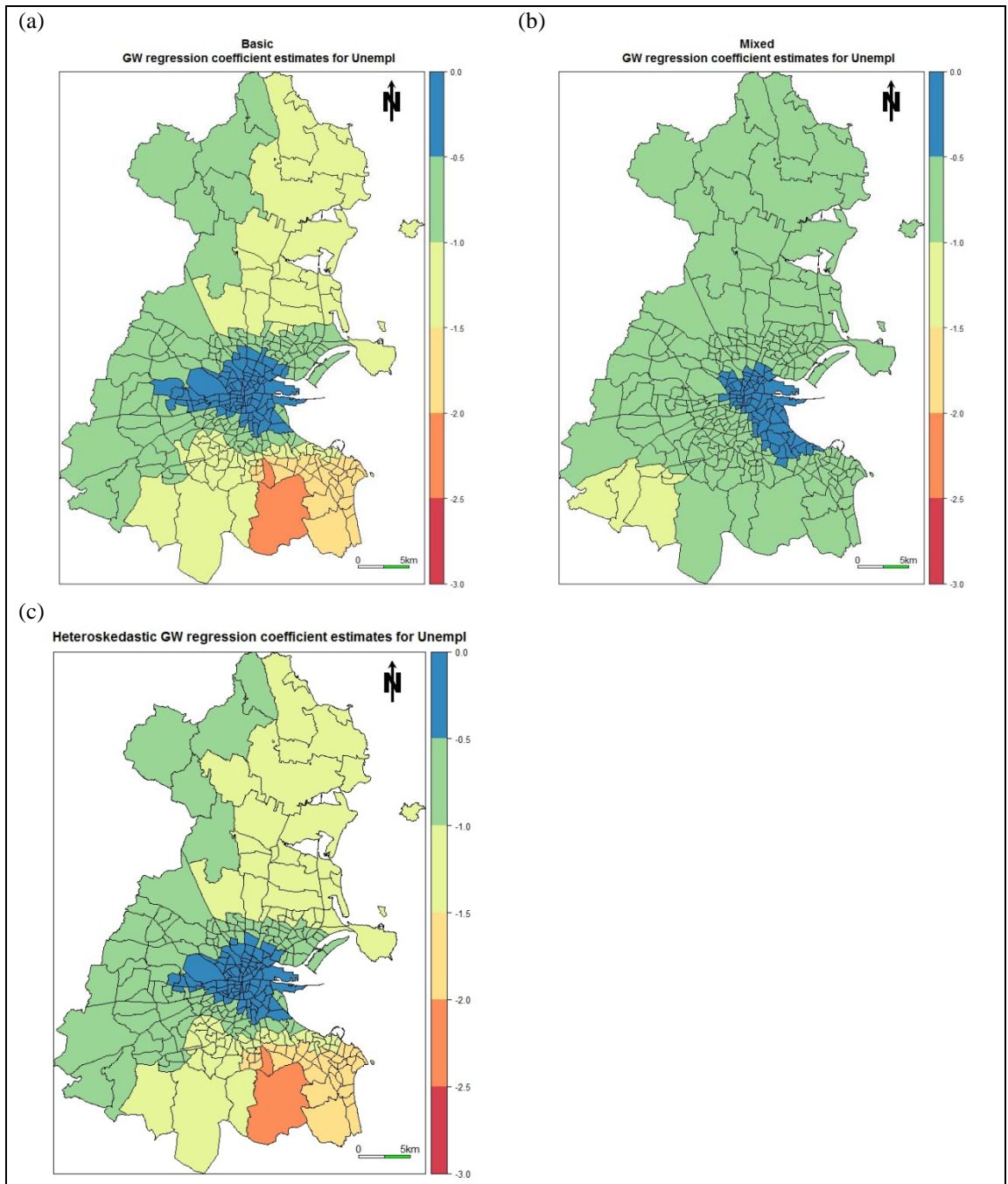

Figure 3 (a) Basic; (b) mixed; and heteroskedastic GW regression coefficients estimates for *Unempl*.

## 5. Further GW regression topics

In this section, we present the use of GW regression functions to conduct: (i) multiple hypothesis tests, (ii) collinearity diagnostics and (iii) heteroskedastic fits.

## 5.1. Multiple hypothesis tests with GW regression

For GW regression, pseudo *t*-values can be used to test, in a purely informal sense, for evidence of local coefficient estimates that are *significantly* different from zero (e.g. *49*). For each coefficient estimate, $\hat{\beta}_k(u,v)$ at location *i*, the pseudo *t*-value can be calculated using:

$$t_{k,i} = \frac{\hat{\beta}_k(u_i, v_i)}{SE(\hat{\beta}_k(u_i, v_i))} \tag{14}$$

where $SE(\hat{\beta}_k(u_i, v_i))$ is the standard error of $\hat{\beta}_k(u_i, v_i)$. For details on the standard error calculations, see (*1*). However, as GW regression yields a separate model at each location *i*, (where each model is calibrated with the same observations but with different weighting schemes), we need to conduct a large number of simultaneous *t*-tests. This operation is likely to result in high order multiple inference problems, where the likelihood of increased type I errors needs to be controlled. In this respect, various standard approaches are available that adjust each test's *p*-value; these include: (a) Benjamini-Hochberg (*36*), (b) Benjamini-Yekutieli (*37*), and (c) Bonferroni (*38*) approaches. All three approaches can be used with GW regression, but a fourth, the Fotheringham-Byrne (*35*) approach is specifically designed for this purpose. This Bonferroni-style adjustment can be briefly described as follows. Let the probability of rejecting one or more true null hypothesis (i.e. the family-wise error rate, FWER) be denoted by $\xi_m$ (with *m* the number of tests). Then the FWER for testing hypotheses about GW regression coefficients can be controlled at $\xi_m$ or less, by selecting

$$\alpha = \xi_m \Big/ \left(1 + p_e - \frac{p_e}{np}\right) \tag{15}$$

where $\alpha$ is the probability of a type I error in the *i*th test; $p_e$ is the effective number of parameters in the GW regression; *np* is the number of parameters in each individual local regression (i.e. the same as that found in the global regression); and *n* is the sample size.

To demonstrate this topic, the results from the basic GW regression of section 4 are investigated. Here all four adjustment approaches are provided in the function ***gwr.t.adjust***, which simply needs to be specified with the results from the GW regression run. Accordingly, we can map the competing test results, where as an example, Figure 4(a) displays the surface of the original (un-adjusted) *p*-values for the *Unempl* coefficient; and Figures 4(b-c) and 5(a-d) present the surfaces for the corresponding adjusted *p*-values (all significant results are coloured red). Observe that the standard (Benjamini-Hochberg, Benjamini-Yekutieli, Bonferroni) approaches adjust the *p*-values to zero or one (i.e. in-significant and significant), whilst the Fotheringham-Byrne approach provides outputs from zero to one. From the five maps, it can be observed that: (1) the un-adjusted *p*-values are similar to that found from Benjamini-Hochberg and Benjamini-Yekutieli adjustments; and (2) the Bonferroni and Fotheringham-Byrne adjustments provide similar results. Commands to conduct all these operations are as follows:

```
R > gwr.t.adj <- gwr.t.adjust(bgwr.res)

R > mypalette.gwr.mht <- brewer.pal(4, "Spectral")

R > X11(width=10,height=12)
R > spplot(gwr.t.adj$SDF, "Unempl_p", key.space = "right", col.regions =
mypalette.gwr.mht, at = c(0, 0.025, 0.05, 0.1, 1.00), main =
"Original p-values for Unempl", sp.layout=map.layout.3)

R > X11(width=10,height=12)
R > spplot(gwr.t.adj$SDF, "Unempl_p_bh", key.space = "right",
col.regions = mypalette.gwr.mht, at = c(0, 0.025, 0.05, 0.1, 1.0000001),
main = "p-values adjusted by Benjamini-Hochberg for Unempl", sp.layout =
map.layout.3)

R > X11(width=10,height=12)
R > spplot(gwr.t.adj$SDF, "Unempl_p_by", key.space = "right",
col.regions = mypalette.gwr.mht, at = c(0, 0.025, 0.05, 0.1, 1.0000001),
main = "p-values adjusted by Benjamini-Yekutieli for Unempl", sp.layout
= map.layout.3)

R > X11(width=10,height=12)
R > spplot(gwr.t.adj$SDF, "Unempl_p_bo", key.space = "right",
col.regions = mypalette.gwr.mht, at = c(0, 0.025, 0.05, 0.1, 1.0000001),
main = "p-values adjusted by Bonferroni for Unempl",
sp.layout=map.layout.3)
```

```
R > X11(width=10,height=12)
R > spplot(gwr.t.adj$SDF, "Unempl_p_fb", key.space = "right",
col.regions = mypalette.gwr.mht, at = c(0, 0.025, 0.05, 0.1, 1.0000001),
main = "p-values adjusted by the Fotheringham-Byrne approach for Unempl",
sp.layout = map.layout.3)
```

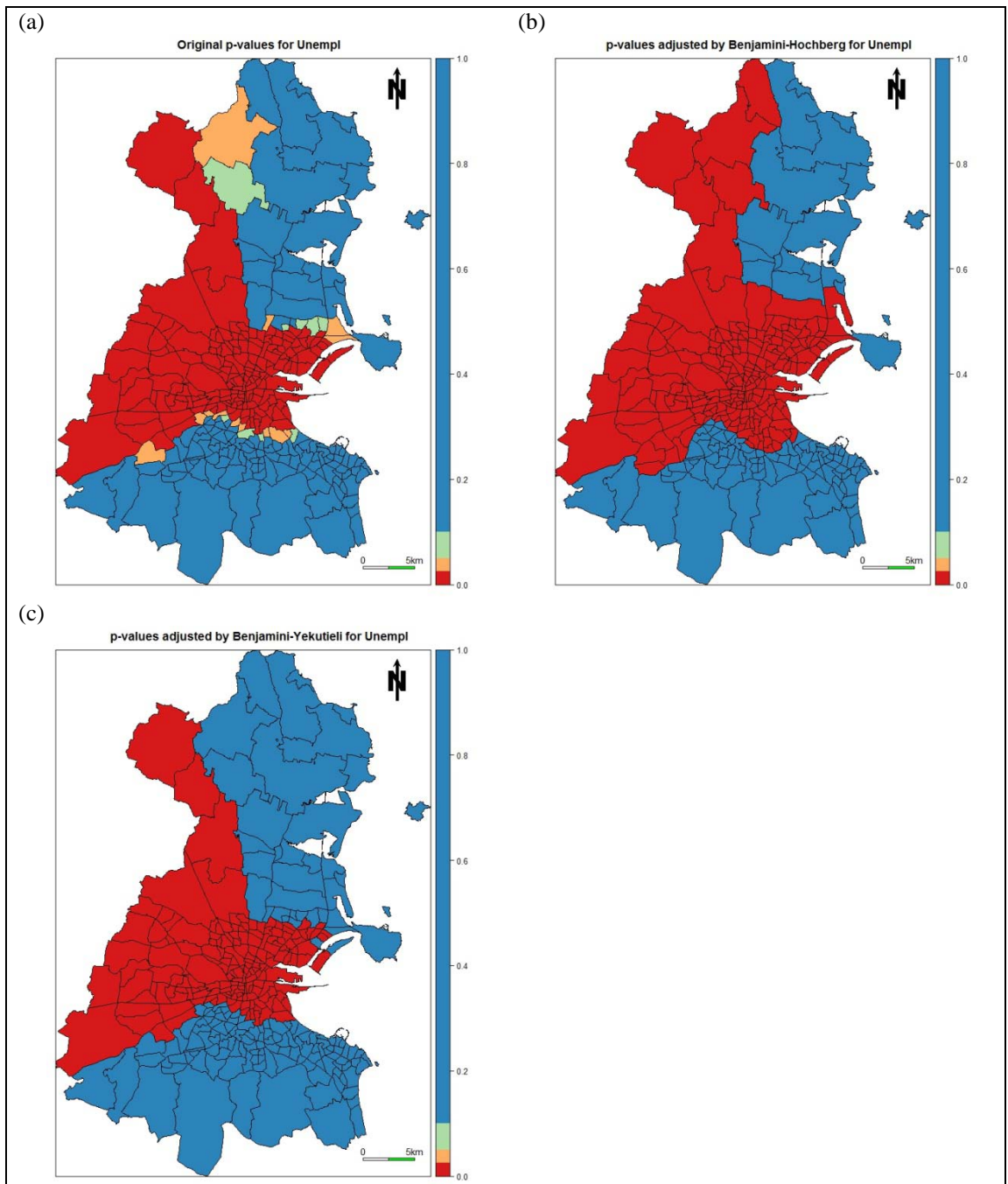

Figure 4 The *p*-values associated with *Unempl* from the basic GW regression of section 4: (a) un-adjusted; (b) adjusted by Benjamini-Hochberg; and (c) adjusted by Benjamini-Yekutieli.

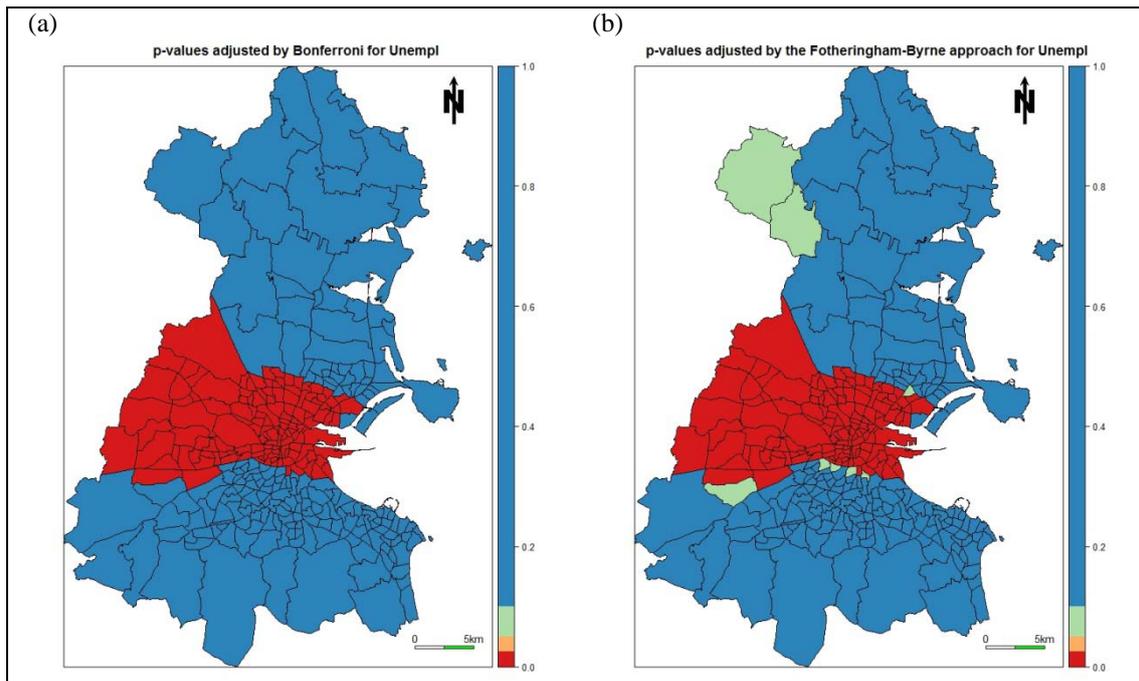

Figure 5 The *p*-values associated with *Unempl* from the basic GW regression of section 4: (a) adjusted by Bonferroni; and (b) adjusted by Fotheringham-Byrne.

## *5.2. Local collinearity diagnostics for a basic GW regression*

The problem of collinearity amongst the predictor variables of a regression model has long been acknowledged and can lead to a loss of precision and power in the coefficient estimates (*42*). This issue is heightened in GW regression since: (A) its effects can be more pronounced with the smaller samples that are used to calibrate each local regression; and (B) if the data is spatially heterogeneous in terms of its correlation structure, some localities may exhibit collinearity when others do not. In both cases, (local) collinearity may cause serious problems in GW regression, when none are found in the corresponding global regression (*6, 29*). Diagnostics to investigate local collinearity in a GW regression model, include finding: (i) local correlations amongst pairs of independent variables; (ii) local variance inflation factors (VIFs) for each independent variable; (iii) local variance decomposition proportions (VDPs); and (iv) local (design matrix) condition numbers (CNs) (see *6, 29*). Accordingly, the following *rules of thumb* can be taken to indicate likely local collinearity problems in the GW regression model: (a) absolute correlation values greater than 0.8 for a given independent variable pair; (b) VIFs greater than 10 for a given independent variable; (c) VDPs greater than 0.5; and (d) CNs greater than 30. Observe that all diagnostics are found at the same spatial scale as each local regression of the GW regression model and can thus be mapped. Observe also that local correlations and local VIFs cannot detect collinearity with the intercept term; thus are considered inferior diagnostics to the combined use of VDPs and CNs (*6*). All four diagnostics are however, considered an integral part of an analytical toolkit that should always be employed in any GW regression analysis. Figure 6a depicts the levels of complexity associated with such investigations (excluding the use of VDPs). Details on the use and merit of these diagnostics, possible model solutions, and critical discussions on this issue with GW regression, can be found in (*6, 23, 29, 39-41, 51, 52*). For this study, our objective is to simply demonstrate some of the collinearity diagnostics used. The application of a (possible) model solution with **GWmodel**, for example, via a locally-compensated ridge GW regression, is demonstrated in Gollini et al. (*23*) using the function *gwr.lcr.* Alternative functions for addressing collinearity in GW regression are found in the **gwrr** R package (*56*).

For a given GW regression specification, local correlations, local VIFs, local VDPs and the local CNs can be found using the function *gwr.collin.diagno.* The same local CNs can also be found using the function *gwr.lcr.* Example maps presenting local correlations, local VIFs and local CNs are given in Figures 6(b-d), reflecting diagnostics for same the basic GW regression of section 4. Scales of each map are chosen to highlight our critical thresholds. Commands to construct these maps are given below. Clearly, significant collinearity is present in our study GW regression model, where DiffAdd appears to be a major cause with respect to its relationship to Age25_44 in central areas of Dublin. As the local CNs are large everywhere, the simple removal of one variable from the analysis may go some way in alleviating this problem; before proceeding to a more locally-focused analysis with some locally-compensated model. This of course brings into question the validity of the results of Kavanagh et al. (*26*), where basic GW regression was applied to this data.

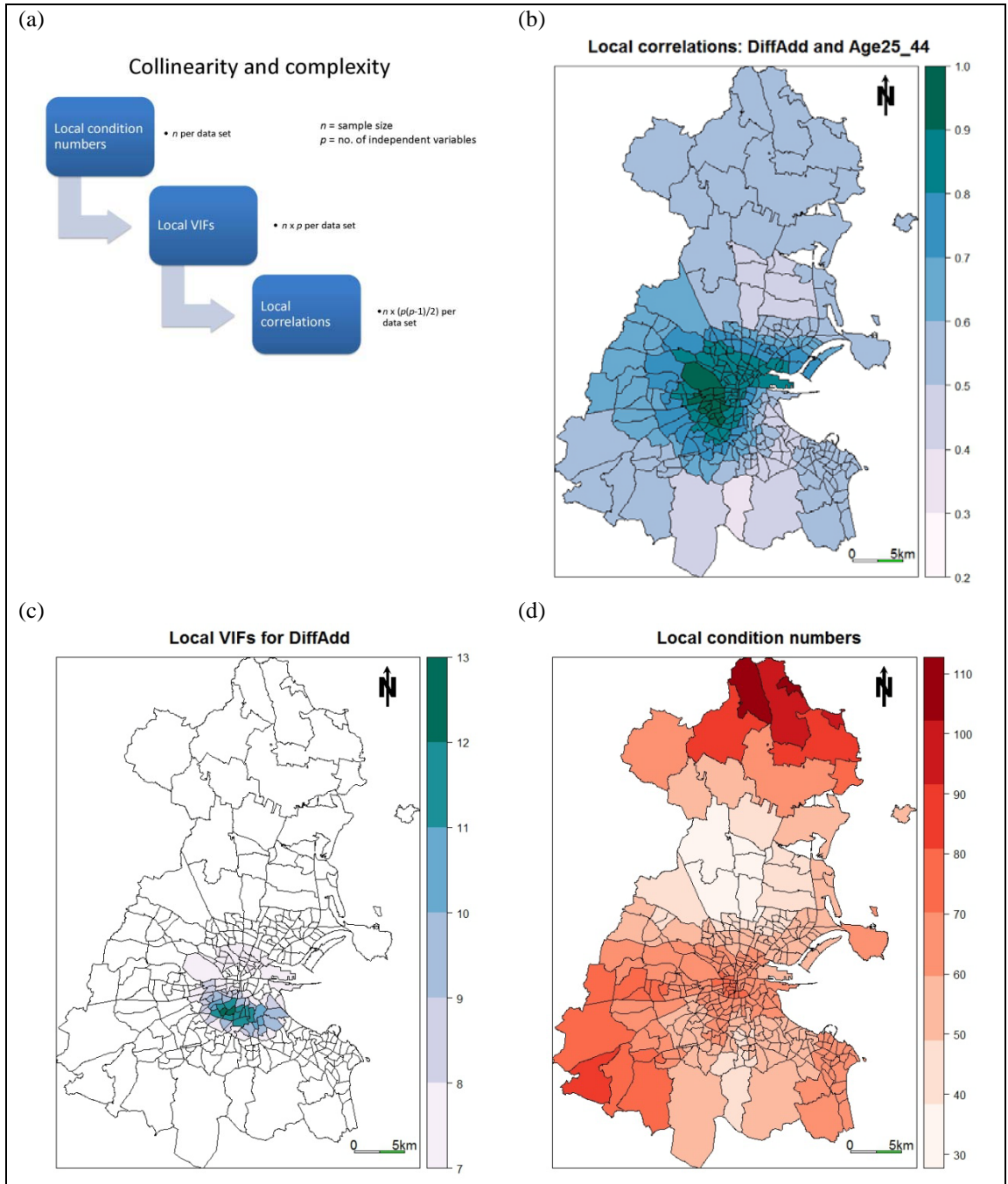

Figure 6 (a) Levels of complexity for different localised collinearity diagnostics; (b) local correlations; (c) local VIFs; and (d) local CNs.

```
R > gwr.coll.data <- gwr.collin.diagno(GenEl2004 ~ DiffAdd + LARent +
SC1 + Unempl + LowEduc + Age18_24 + Age25_44 + Age45_64,
data = Dub.voter, bw = bw.gwr.1, kernel = "bisquare", adaptive = TRUE)
```

```
R > mypalette.coll.1 <-brewer.pal(8,"PuBuGn")
R > X11(width=10,height=12)
R > spplot(gwr.coll.data$SDF,"Corr_DiffAdd.Age25_44",key.space = "right",
col.regions=mypalette.coll.1,at=c(0.2,0.3,0.4,0.5,0.6,0.7,0.8,0.9,1),
par.settings=list(fontsize=list(text=15)),
main=list(label="Local correlations: DiffAdd and Age25_44", cex=1.25),
sp.layout=map.layout.3)

R > mypalette.coll.2 <-brewer.pal(6,"PuBuGn")
R > X11(width=10,height=12)
R > spplot(gwr.coll.data$SDF,"DiffAdd_VIF",key.space = "right",
col.regions=mypalette.coll.2,at=c(7,8,9,10,11,12,13),
par.settings=list(fontsize=list(text=15)),
main=list(label="Local VIFs for DiffAdd", cex=1.25),
sp.layout=map.layout.3)

R > mypalette.coll.3 <-brewer.pal(8,"Reds")
R > X11(width=10,height=12)
R > spplot(gwr.coll.data$SDF,"local_CN",key.space = "right",
col.regions=mypalette.coll.3,cuts=7,
par.settings=list(fontsize=list(text=15)),
main=list(label="Local condition numbers", cex=1.25),
sp.layout=map.layout.3)
```

### *5.3. Heteroskedastic GW regression*

Basic GW regression assumes that the error term is normally distributed with zero mean and constant (stationary) variance over the study region ($\varepsilon_i \sim N(0, \sigma^2)$). An extension of GW regression is possible, which allows a non-stationary error variance ($\varepsilon_i \sim N(0, \sigma^2(u_i, v_i))$). This (possibly more realistic) model was first proposed in Fotheringham et al. (*1*) and has been further extended to a predictive form in Harris et al. (*18*). Details of such heteroskedastic GW regressions can be found in (*1,18*), where an iterative modelling technique is used, requiring the model to converge to some pre-specified level of tolerance. An alternative (parametric) heteroskedastic GW regression can be found the work of Paez et al. (*54*).

For GWmodel, the function ***gwr.hetero*** allows the non-parametric version to be specified. Currently, the model is only given in a very rudimentary form, where the kernel function used to control the coefficient estimates is also used to control the local error variances; which are themselves approximated by the squared residuals ($e^2$). Outputs from *gwr.hetero* are the regression coefficients only, which can be compared to the corresponding coefficients found using a basic GW regression (*gwr.basic*). As an example, Figure 3(c) displays the coefficient surface for *Unempl* from the heteroskedastic model. Clearly, there is little difference from the coefficient surface of the basic model (Figure 3(a)). Thus modelling with a stationary error variance appears reasonable. The use of a non-stationary error variance can be useful, however, for improved measures of uncertainty (*18*) and for outlier detection (*49*). Commands to fit our heteroskedastic model are as follows, noting that the function *gwr.hetero* is specified with the same bandwidth, as that found for the basic model.

```
R > hgwr.res <- gwr.hetero(GenEl2004 ~ DiffAdd + LARent + SC1 + Unempl
+ LowEduc + Age18_24 + Age25_44 + Age45_64, data = Dub.voter, bw =
bw.gwr.1, kernel = "bisquare", adaptive = TRUE)

R > X11(width=10,height=12)
R > spplot(hgwr.res, "Unempl", key.space = "right", col.regions =
mypalette.gwr, at = c(-3, -2.5, -2, -1.5, -1, -0.5, 0), main =
"Heteroskedastic GW regression coefficient estimates for Unempl",
sp.layout=map.layout.3)
```

## 6. GW Discriminant Analysis

Discriminant analysis (DA) allows the modelling and prediction of a categorical dependent variable explained by a set of independent variables. As with GW regression, the relationships between the dependent and independent variables may vary across space. In such a cases, a GW discriminant analysis (GW DA) (*15*) provides a useful investigative tool, where the discrimination rule is localised. DA (and in turn, GW DA) provides an alternative to logistic regression (and in turn, GW logistic regression); a useful comparison of which can be found in the simulation study of Pohar et al. (*43*), where guidelines to choosing one method in preference to the other are presented.

*6.1. GW discriminant analysis*

The theoretical context for DA is briefly described. Suppose a population, of which each object belongs to $k$ possible categories; and a training set $X$, where each row vector $x_i$ indicates an observation belonging to category $l$ ($l \in \{1,\cdots,k\}$); then for an observation vector $x$, the discrimination rule is to assign $x$ to the $l$th category with the maximum probability that $x$ belongs to this category, say

$$P_l(x) = \max_{j\in\{1,\cdots,k\}}\{P_j(x) = p_j f_j(x)\}. \qquad (16)$$

Here $p_j f_j(x)$ is proportional to the posterior probability of an observation arising from population $j$ once the value of $x$ is known. Now a linear DA (LDA) assumes that the distribution for each category is multivariate normal with a discrimination rule of

$$P_l(x) = \max_{j\in\{1,\cdots,k\}}\left\{P_j(x) = p_j \frac{1}{(2\pi|\Sigma|)^{q/2}} e^{-\frac{1}{2}(x-\mu_j)^T \Sigma^{-1}(x-\mu_j)}\right\} \qquad (17)$$

where $\Sigma$ is the covariance matrix, $q$ is the number of independent variables in $x$, and $\mu_j$ is the mean for population $j$. This function can be simplified by taking logs and changing signs

$$LP_l(x) = \min_{j\in\{1,\cdots,k\}}\left\{LP_j(x) = \tfrac{1}{2}(x-\mu_j)^T \Sigma^{-1}(x-\mu_j) + \tfrac{q}{2}\log(|\Sigma|) - \log(p_j)\right\}. \qquad (18)$$

LDA assumes that $\Sigma$ is identical for each category, whilst an alternative, quadratic DA (QDA) assumes that $\Sigma$ is different for each category $j$, where its discrimination rule replaces $\Sigma$ in equation (18) with $\Sigma_j$.

GW DA is a direct local adaption of DA, with the chosen discrimination rule varying across space. Here the stationary mean and covariance estimates of DA are replaced with respective GW mean estimates (equation (1)) and GW covariance estimates (equation (4)) in the discrimination rules for both LDA and QDA. Thus a local LDA or QDA can be found at any location $(\boldsymbol{u}, \boldsymbol{v})$ using GW DA. Bandwidth selection follows a cross-validation approach, where an optimum bandwidth is identified by minimising this score:

$$CV_{GWDA}(b) = \sum_i P_{\neq i}(b) \quad (19)$$

where $P_{\neq i}$ is the proportion of incorrect assignments when the observation *i* is removed from the sample data.

## 6.2. Example: GW discriminant analysis

To demonstrate a GW DA, we use the `USelect` data described in section 2. Here we calibrate a GW DA using the function *gwda* in **GWmodel**, together with a standard DA (LDA) using the function *lda* from the **MASS** R package (*44*). The GW DA is conducted with an adaptive bandwidth (bi-square kernel) with its optimum found using the function *bw.gwda*. The resultant confusion matrices are presented in Table 3. Here, the DA classification accuracy is 72.5%, whilst GWDA provides a slightly improved classification accuracy of 74.0%. An interesting feature is that the global model predicts only one county in the 'Borderline' category. Results of the actual presidential election are mapped at the county level in Figure 7(a), where Bush was a clear winner in most counties, while the election was more competitive in areas like Wisconsin and Maine. The classification results using DA and GW DA are mapped in Figures 7(b-c), where the spatial pattern in the GW DA classifications appears marginally closer to the true results than that found with the DA. Commands to conduct all these operations and visualisations are given below.

Table 3 DA and GW DA confusion matrices.

|  | **Borderline** | **Bush** | **Kerry** | **Total** |
|---|---:|---:|---:|---:|
| *DA:* | | | | |
| **Borderline** | 1 | 1 | 4 | 6 |
| **Bush** | 543 | 2099 | 166 | 2808 |
| **Kerry** | 92 | 49 | 156 | 297 |
| **Total** | 636 | 2149 | 326 | 3111 |
| *GW DA:* | | | | |
| **Borderline** | 29 | 15 | 12 | 56 |
| **Bush** | 522 | 2103 | 145 | 2770 |
| **Kerry** | 85 | 31 | 169 | 285 |
| **Total** | 636 | 2149 | 326 | 3111 |

```
R > library(MASS)
R > data(USelect)

R > lda.res <-
lda(winner~unemploy+pctcoled+PEROVER65+pcturban+WHITE,USelect2004)
R > lda.pred <- predict(lda.res, USelect2004)
R > lda.SDF <- SpatialPolygonsDataFrame(Sr = polygons(USelect2004),
data = data.frame(lda.pred), match.ID = F)
R > CM.lda <- confusion.matrix(USelect2004$winner, lda.pred$class)
R > CM.lda
R > lda.cr <- length(which(USelect2004$winner ==
lda.pred$class))/nrow(USelect2004@data)
R > lda.cr

R > Dmat <- gw.dist (dp.locat = coordinates (USelect2004))

R > bw.gwda.ab <-
bw.gwda(winner~unemploy+pctcoled+PEROVER65+pcturban+WHITE,
USelect2004, kernel= "bisquare", adaptive=T, dMat=Dmat)
R > gwda.ab <- gwda(winner~unemploy+pctcoled+PEROVER65+pcturban+WHITE,
USelect2004, bw=bw.gwda.ab, kernel= "bisquare", adaptive=T, dMat=Dmat)
R > print(gwda.ab)
R > CM.gwda.ab <- confusion.matrix(USelect2004$winner,
gwda.ab$SDF$group.predicted)
R > CM.gwda.ab
R > gwda.cr <- length(which(USelect2004$winner ==
gwda.ab$SDF$group.predicted))/nrow(USelect2004@data)
R > gwda.cr

R > mypalette.gwda <- brewer.pal(3, "Spectral")
R > xy <- coordinates(USelect2004)

R > X11(width=16,height=8)
R > USelect2004$winner <- factor(USelect2004$winner, levels =
c("Borderline","Bush", "Kerry"))
R > spplot(USelect2004, "winner", key.space = "right",
col.regions = mypalette.gwda,
par.settings=list(fontsize=list(text=20)),
main = list(label="Results of the 2004 US presidential election",
cex=1.25))
```

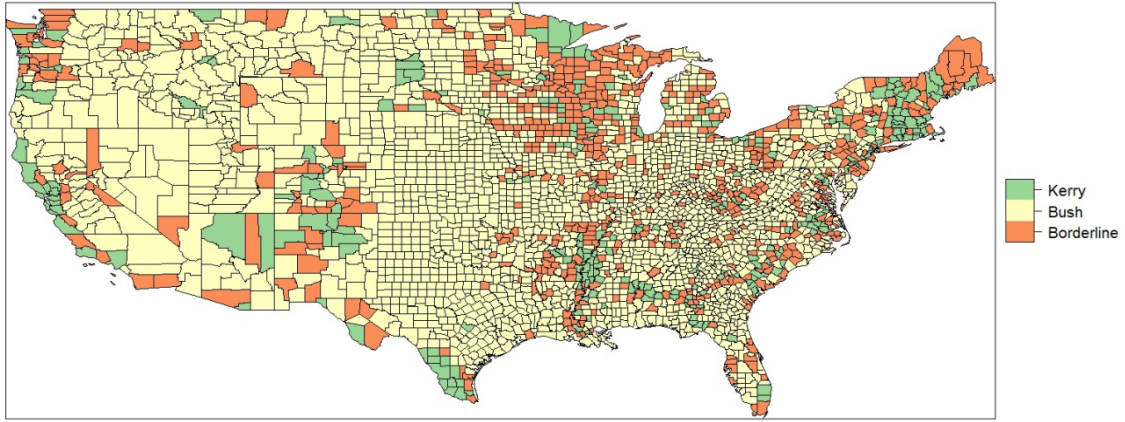

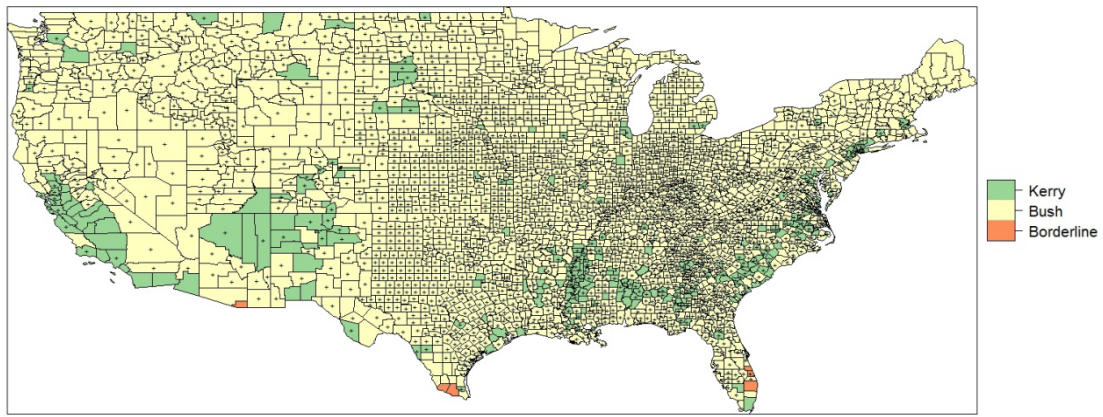

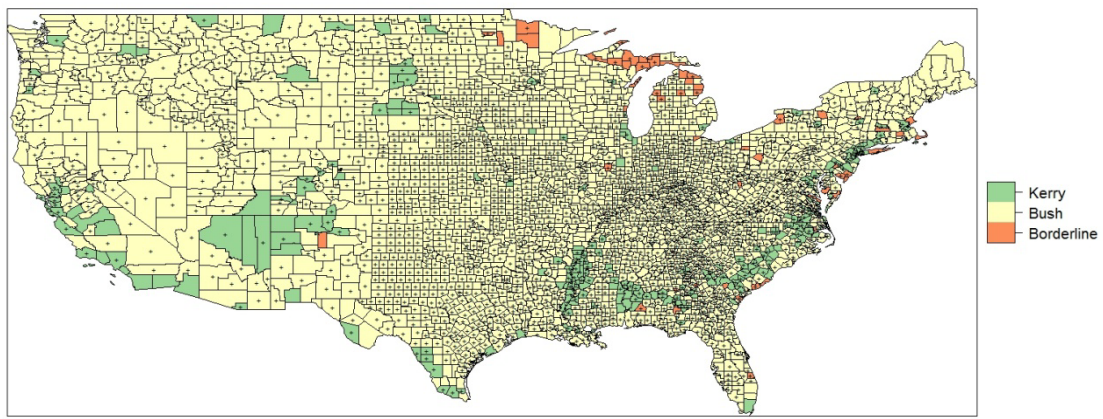

Figure 7 (a) Results of 2004 US presidential election. Classification results using: (b) DA; and (c) GW DA.

```
R > X11(width=16,height=8)
R > pts.lda.correct <- list("sp.points", xy[which(USelect2004$winner ==
lda.pred$class),],  cex=1, pch="+", col="black")
R > lda.SDF$class <- factor(lda.SDF$class,
levels = c("Borderline","Bush", "Kerry"))
R > spplot(lda.SDF, "class", key.space = "right",
col.regions = mypalette.gwda,
par.settings=list(fontsize=list(text=20)),
main = list(label="Classification results using DA ", cex=1.25),
sub=list(label="+ the correct classification", cex=1),
sp.layout=pts.lda.correct)

R > X11(width=16,height=8)
R > pts.gwda.ab.correct <- list("sp.points", xy[which(USelect2004$winner
== gwda.ab$SDF$group.predicted),],  cex=1, pch="+", col="black")
R > gwda.ab$SDF$group.predicted <- factor(gwda.ab$SDF$group.predicted,
levels = c("Borderline","Bush", "Kerry"))
R > spplot(gwda.ab$SDF, "group.predicted", key.space = "right",
col.regions = mypalette.gwda,
par.settings=list(fontsize=list(text=20)),
main = list(label="Classification results using GW DA", cex=1.25),
sub=list(label="+ the correct classification", cex=1),
sp.layout=pts.gwda.ab.correct)
```

## 7. Enhanced kernel bandwidth selection

In **GWmodel**, a number of functions are provided to aid bandwidth selection. These include: ***bw.ggwr***, ***bw.gwda***, ***bw.gwpca***, ***bw.gwr*** and ***bw.gwr.lcr*** for automatic bandwidth selection when calibrating a generalised GW regression, a GW DA, a GW PCA, basic GW regression and GW regression with a local compensated ridge term, respectively. However, it is not always recommended to simply plug the resultant (optimal) bandwidth into the given GW model, without first checking the behaviour of the full bandwidth function. Here we demonstrate how to: (i) investigate for multiple minima in this function and (ii) assess if (outlying) observations adversely affect the behaviour of this function. In this respect, **GWmodel** provides the following functions for constructing a cross-validation (CV) bandwidth function (i.e. bandwidth vs. the CV score): ***ggwr.cv***, ***gwpca.cv***, ***gwr.cv*** and ***gwr.lcr.cv*** (with ***gwda.cv*** still to be coded). Similarly, **GWmodel** provides the following functions for finding the CV score data at each observation location, for a given bandwidth*:*

***ggwr.cv.contrib***, ***gwpca.cv.contrib***, ***gwr.cv.contrib*** and ***gwr.lcr.cv.contrib*** (with ***gwda.cv.contrib*** still to be coded). Observe that the CV score data is summed to provide the CV score for a given bandwidth. As an example of using these functions, we further investigate the GW PCA conducted in section 3. Here we use the ***bw.gwpca***, *gwpca.cv* and *gwpca.cv.contrib* functions. Thus the CV bandwidth function, and a histogram and map of the CV score data for an optimal bandwidth of $N = 131$, are found as follows (and presented in Figures 8(a-c)).

```
R > library(classInt)

R > sample.n <- 322
R > bwd.range.adapt <- c(seq(40,sample.n,by=20))
R > cv.score <- matrix(nrow=length(bwd.range.adapt),ncol=1)
R > for(i in 1:length(bwd.range.adapt)) cv.score[i] <-
gwpca.cv(bwd.range.adapt[i],Data.scaled,Coords,k=3,robust=F,
kernel="bisquare",adaptive=TRUE,p=2,theta=0,longlat=F)

R > X11(width=6,height=6)
R > plot(bwd.range.adapt,cv.score,ylab="",xlab="",cex=1,pch=19)
R > title(ylab = list("CV score", cex=1.25, col="black", font=1))
R > title(xlab = list("No. of nearest neighbours", cex=1.25, col="black",
font=1))
R > title(main = list("GW PCA: Bandwidth function", cex=1.5, col="black",
font=1))

R > cv.score.data.opt <-
gwpca.cv.contrib(Data.scaled,Coords,bw=131,k=3,robust=F,kernel="bisquare",a
daptive=TRUE,p=2,theta=0,longlat=F)

R > X11(width=6,height=6)
R > hist(cv.score.data.opt,ylab="",xlab="",main="")
R > title(ylab = list("Frequency", cex=1.25, col="black", font=1))
R > title(xlab = list("CV score data", cex=1.25, col="black", font=1))
R > title(main = list("GW PCA: CV score data for a bandwidth of 131",
cex=1.5, col="black", font=1))

R > Dub.voter$cv.score.data.opt <- cv.score.data.opt
R > mypalette.cv.score <- brewer.pal(9,"Spectral")
R > b1 <- classIntervals(cv.score.data.opt,n=9,style="quantile")
b1$brks[length(b1$brks)] <- b1$brks[length(b1$brks)] * 1.001

R > X11(width=10,height=12)
R > spplot(Dub.voter,"cv.score.data.opt",key.space = "right",
col.regions=mypalette.cv.score, cuts=8,
par.settings=list(fontsize=list(text=15)),
main=list(label="GW PCA: CV score data for a bandwidth of 131", cex=1.25),
sub=list(label="Quantile intervals",cex=1.05),
sp.layout=map.layout.3)
```

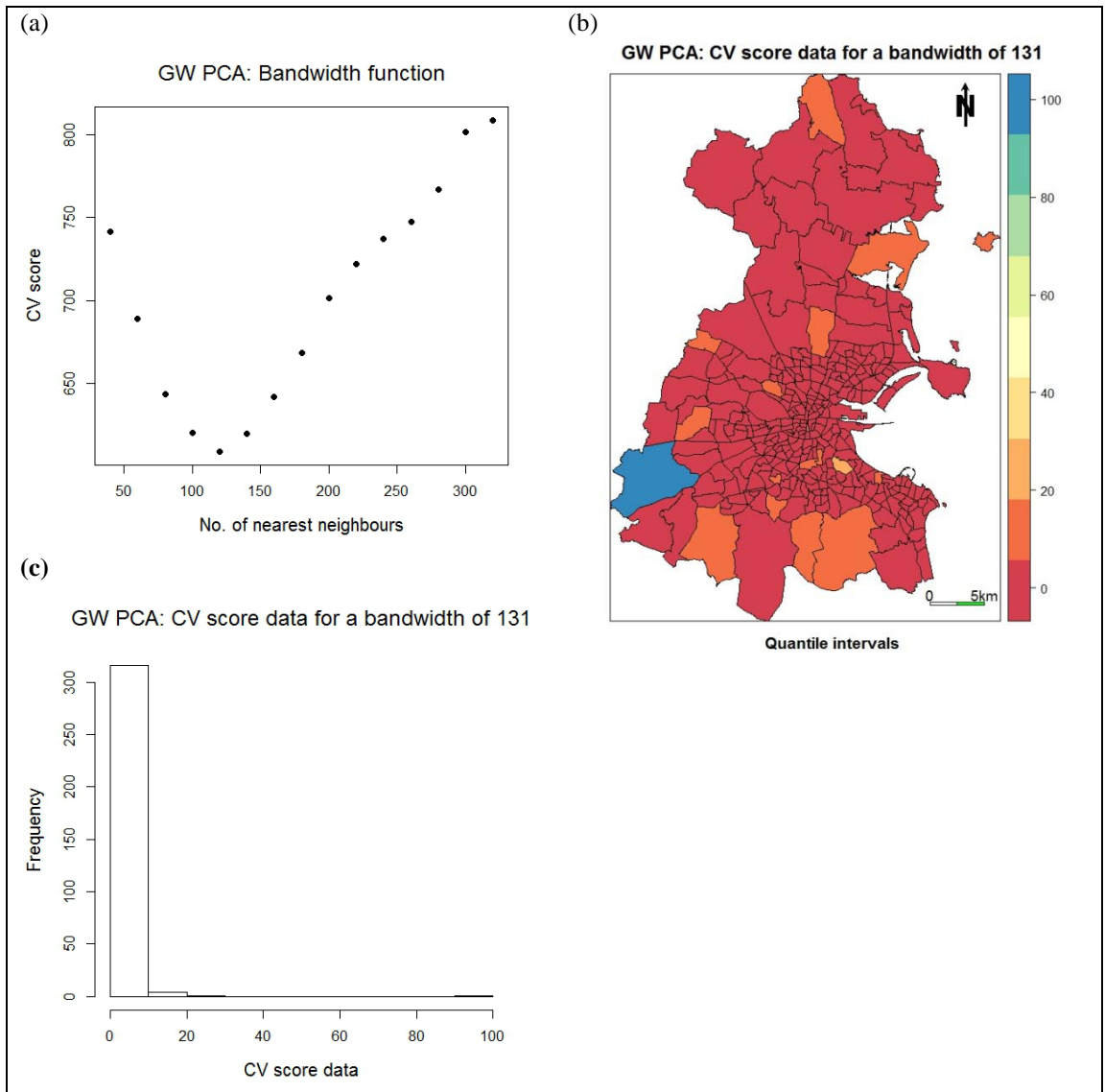

Figure 8 GW PCA calibration: (a) adaptive CV bandwidth function; (b) map and (c) histogram of the CV score data for the optimal bandwidth of $N = 131$.

It is clear from Figures 8(a-c), that the CV bandwidth function is well-behaved, reaching a clear minimum at $N = 131$; and thus provides re-assurance in this bandwidth's use. At this specific bandwidth, the CV score data is heavily positively skewed, with one extreme value at 98.4 that corresponds to an ED in the south-west of Greater Dublin. The census data at this ED warrants additional scrutiny and maybe in error.

# 8. Concluding remarks

This study, together with its companion study (*23*), demonstrates the application of a wide range of techniques for investigating spatial heterogeneity, using functions provided by the **GWmodel** R package. Topics include that of (i) GW summary statistics, (ii) GW principal component analysis, (iii) GW regression, and (iv) GW Discriminant Analysis. The GW modelling paradigm provides a simple, yet powerful analytical toolkit for exploring change in a statistical model's parameters and outputs across space; a paradigm that continues to evolve (e.g. *16, 21, 24, 45, 46, 47, 48, 52, 53*). Functions for these more recent advances in GW modelling will be incorporated into **GWmodel** in due course.

**Acknowledgements**

Research presented in this paper was funded by a Strategic Research Cluster grant (07/SRC/I1168) by Science Foundation Ireland under the National Development Plan. The authors gratefully acknowledge this support.